\documentclass[aps,twocolumn,footinbib,floatfix]{revtex4}

\usepackage{amssymb}
\usepackage[pdftex]{graphicx}	
	\DeclareGraphicsExtensions{.pdf,.png,.jpg}
\usepackage[applemac]{inputenc}
\usepackage{amsmath,amssymb,amsthm}
\usepackage{mathbbol,bbm}
\usepackage{graphicx,float}
\usepackage[final]{showkeys}
\usepackage{bm,dsfont}
\usepackage{color}

\newcommand{\A}{\mathrm{A}}
\newcommand{\B}{\mathrm{B}}
\newcommand{\rmS}{\mathrm{S}}

\newcommand{\calH}{\mathcal{H}}

\def\ket#1{\left|#1\right>}
\def\bra#1{\left<#1\right|}
\def\Tr{ {\rm{Tr }}}

 \def\ee{\mathord{\rm e}}
 
 \def\ii{\mathord{\rm i}}

\renewcommand{\ii}{{\rm i}}
\renewcommand{\ee}{{\rm e}}

\usepackage{booktabs}

\begin{document}
%\title{Quantifier for Spatial Correlations of General Quantum Dynamics}
%\title{Quantifying Correlations of General Quantum Dynamics}
\title{Quantifying Spatial Correlations of General Quantum Dynamics}
\author{\'Angel Rivas and Markus M\"uller}
\affiliation{Departamento de F\'{\i}sica Te\'orica I, Universidad Complutense, 28040 Madrid, Spain}

\vspace{-3.5cm}

\begin{abstract}
Understanding the role of correlations in quantum systems is both a fundamental challenge as well as of high practical relevance for the control of multi-particle quantum systems. Whereas a lot of research has been devoted to study the various types of correlations that can be present in the states of quantum systems, in this work we introduce a general and rigorous method to quantify the amount of correlations in the dynamics of quantum systems. Using a resource-theoretical approach, we introduce a suitable quantifier and characterize the properties of correlated dynamics. Furthermore, we benchmark our method by applying it to the paradigmatic case of two atoms weakly coupled to the electromagnetic radiation field, and illustrate its potential use to detect and assess spatial noise correlations in quantum computing architectures.
\end{abstract}


\maketitle

% 03.65.Ca 	Formalism
% 03.65.Ta 	Foundations of quantum mechanics; measurement theory (for optical tests of quantum theory, see 42.50.Xa)
% 03.67.Mn	Entanglement measures, witnesses, and other characterizations
% 03.65.Ud  Entanglement and quantum nonlocality
% 67.85.-d	Ultracold gases, trapped gases
% 03.65.Yz	Decoherence; open systems; quantum statistical methods
% 42.50.Lc 	Quantum fluctuations, quantum noise, and quantum jumps

%%%%%%%%%%%%
\section{Introduction}
%%%%%%%%%%%%

Quantum systems can display a wide variety of dynamical behaviors, in particular depending on how the system is affected by its coupling to the surrounding environment. One interesting feature which has attracted much attention is the presence of memory effects (non-Markovianity) in the time evolution. These typically arise for strong enough coupling between the system and its environment, or when the environment is structured, such that the assumptions of the well-known weak-coupling limit \cite{BrPe02,GardinerZoller04,Libro} are no longer valid. Whereas memory effects (or time correlations) can be present in any quantum system exposed to noise, another extremely relevant feature, which we will focus on in this work, are correlations in the dynamics of different parts of multi-partite quantum systems. Since different parties of a partition are commonly, though not always, identified with different places in space, without loss of generality we will in the following refer to these correlations between different subsystems of a larger system as spatial correlations.

Spatial correlations in the dynamics give rise to a wide plethora of interesting phenomena ranging from super-radiance \cite{Dicke} and super-decoherence \cite{14-Qubit} to sub-radiance \cite{Pillet} and decoherence-free subspaces \cite{Zanardi,Lidar1,Lidar2,Wineland,Haeffner}. Moreover, clarifying the role of spatial correlations in the performance of a large variety of quantum processes, such as e.g. quantum error correction \cite{Clemmens,Klesse,Kitaev,Preskill2013,Novais,Shabani}, photosynthesis and excitation transfer \cite{Caruso,Aspuru,Nazir,Olaya,Nalbach,Silbey,Olbrich,Sarovar,Schulten,Mukamel,Jeske3}, dissipative phase transitions \cite{Diehl,Verstraete,Igor,Lee,Schindler} and quantum metrology \cite{Jeske2} has been and still is an active area of research.

Along the last few years, numerous works have aimed at quantifying up to which extent quantum dynamics deviates from the Markovian behavior, see e.g. \cite{Wolf,BrLaPi,RHP,Sun,Mutual,Mauro,MichaelHall,Bylicka,Review}. However, much less attention has been paid to develop quantifiers of spatial correlations in the dynamics, although some works e.g. \cite{Jeske1,Joe} have addressed this issue for some specific models. This may be partially due to the well-known fact that under many, though not all practical circumstances, dynamical correlations can be detected by studying the time evolution of correlation functions of properly chosen observables $\mathcal{O}_A$ and $\mathcal{O}_B$, acting respectively on the two parties A and B of interest. For instance, in the context of quantum computing, sophisticated methods to witness the correlated character of quantum dynamics, have been developed and implemented in the laboratory \cite{Joe}. Indeed, any correlation $C(\mathcal{O}_A, \mathcal{O}_B)=\langle \mathcal{O}_A \otimes  \mathcal{O}_B \rangle-\langle \mathcal{O}_A \rangle \langle \mathcal{O}_B \rangle$ detected during the time evolution of an initial product state, $\rho = \rho_A \otimes \rho_B$, witnesses the correlated character of the dynamics. However, note that there exist highly correlated dynamics, which cannot be realized by a combination of local processes, which do not generate any such correlation,
e.g.~the swap process between two parties. Such dynamics can either act on internal degrees of freedom, induced e.g.~by the action of a swap gate acting on two qubits \cite{NC00}, or can correspond to (unwanted) external dynamics, caused e.g.~by correlated hopping of atoms in an optical lattice \cite{Folling, Lewenstein-book} or crystal melting and subsequent recooling dynamics in trapped-ion architectures \cite{Naegerl}.

Thus, it is of eminent importance to develop methods which allow us to detect the presence or absence of spatial correlations in the dynamics, without a priori knowledge of the underlying microscopic dynamics, and do not require us to resort to adequately chosen ``test'' observables and initial ``test'' quantum states. Such methods should furthermore provide a rigorous ground to quantitatively compare the amount of spatial correlations in different dynamical processes. These characteristics are essential for a ``good'' correlation quantifier that can be used to study spatial correlations in quantum dynamics from a fundamental point of view \cite{Kraus, Nielsen-Dawson, Linden}, to clarify their role in physical processes \cite{Clemmens,Klesse,Kitaev,Preskill2013,Novais,Shabani,Caruso,Aspuru,Nazir,Olaya,Nalbach,Silbey,Olbrich,Sarovar,Schulten,Mukamel,Jeske3,Diehl,Verstraete,Igor,Lee,Schindler,Jeske2}, as well as to measure and quantify spatial correlations in the dynamics of experimental quantum systems.

It is the aim of this work to introduce a method to quantify the degree of correlation in general quantum dynamics from a fundamental view point. Specifically,

\smallskip
\noindent i) we propose a theoretical framework and formulate a general measure to assess the amount of spatial correlations of quantum dynamics without resorting to any specific physical model. To this end, we adopt a resource theory approach, and formulate a fundamental law that any faithful measure must satisfy.

\smallskip
\noindent ii) Within this framework, we study the properties that a dynamics has to fulfill to be considered as maximally correlated.

\smallskip
\noindent iii) We apply our measure to the paradigmatic quantum-optical model of two two-level atoms radiating into the electromagnetic vacuum. This case exemplifies the working principle of our measure and quantitatively confirms the expectation that spatial dynamical correlations decay with increasing interatomic distance and for long times.

\smallskip
\noindent iv) Finally, we illustrate this formalism with a second example in the context of quantum computing, where quantum error correction protocols rely on certain assumptions on (typically sufficiently small) noise strengths and noise correlations. Specifically, we consider two qubits subject to local thermal baths that suffer some residual interaction which induces a correlated noisy dynamics. Our method reveals the remarkable fact that, under keeping the overall error probability for the two qubits constant, the degree of spatial correlations decays very rapidly as the bath temperature increases. This suggests that, in some situations, noise addition as e.g.~by a moderate increase of the environmental temperature, can be beneficial to tailor specific desired noise characteristics.

%%%%%%%%%%%%
\section{Measure of Correlations for Dynamics}
%%%%%%%%%%%%

%%%%%%%%%%%%
\subsection{Uncorrelated Dynamics}
%%%%%%%%%%%%
Let us consider a bipartite quantum system $\rmS=\A\B$ undergoing some dynamics given by a completely positive and trace preserving (CPT) map $\mathcal{E}_\rmS$ [without loss of generality we shall assume $\dim(\mathcal{H}_\A)=\dim(\mathcal{H}_\B)=d$ and so $d_{\rmS}:=\dim(\mathcal{H}_\rmS)=d^2$]. This dynamics is said to be uncorrelated with respect to the subsystems $\A$ and $\B$ if it can be decomposed as $\mathcal{E}_\rmS=\mathcal{E}_\A\otimes\mathcal{E}_\B$, with CPT maps $\mathcal{E}_\A$ and $\mathcal{E}_\B$ acting on $\A$ and $\B$, respectively. Otherwise it is said to be correlated.

The central tool of our construction is the Choi-Jamio{\l}kowski isomorphism \cite{Choi,Jamiolkowski}, which provides a one-to-one map of a given quantum dynamics to an equivalent representation in the form of a quantum state in an enlarged Hilbert space. This mapping allows us to use tools developed for the quantification of correlations in quantum states for our purpose of quantifying correlations in quantum dynamics. Thus, consider a second $d^2-$dimensional bipartite system $\rmS'=\A'\B'$, and let $\ket{\Phi_{\rmS\rmS'}}$ be the maximally entangled state between $\rmS$ and $\rmS'$,
\begin{equation}\label{MaxEnta}
\ket{\Phi_{\rmS\rmS'}}:=\frac{1}{d}\sum_{j=1}^{d^2}\ket{jj}_{\rmS\rmS'}=\frac{1}{d}\sum_{k,\ell=1}^d\ket{ k \ell}_{\A\B}\otimes \ket{k\ell}_{\A'\B'}.
\end{equation}
Here, $\ket{j}$ denotes the state vector with 1 at the $j$-th position and zero elsewhere (canonical basis). The Choi-Jamio{\l}kowki representation of some CPT map $\mathcal{E}_\rmS$ on $\rmS$ is given by the $d^4-$dimensional state
\begin{equation}\label{CJstate}
\rho^{\rm CJ}_\rmS:=\mathcal{E}_\rmS\otimes\mathds{1}_{\rmS'}(\ket{\Phi_{\rmS\rmS'}}\bra{\Phi_{\rmS\rmS'}}),
\end{equation}
where $\mathds{1}_{\rmS'}$ denotes the identity map acting on $\rmS'$. The entire information about the dynamical process $\mathcal{E}_\rmS$ is contained in this unique state.

%%%%%%%%%%%%%%%%%%%%%%%%%%
%%%%%%%%%%%%%%%%%%%%%%%%%%
\begin{figure}[t]
\begin{center}
\includegraphics[width=1\columnwidth]{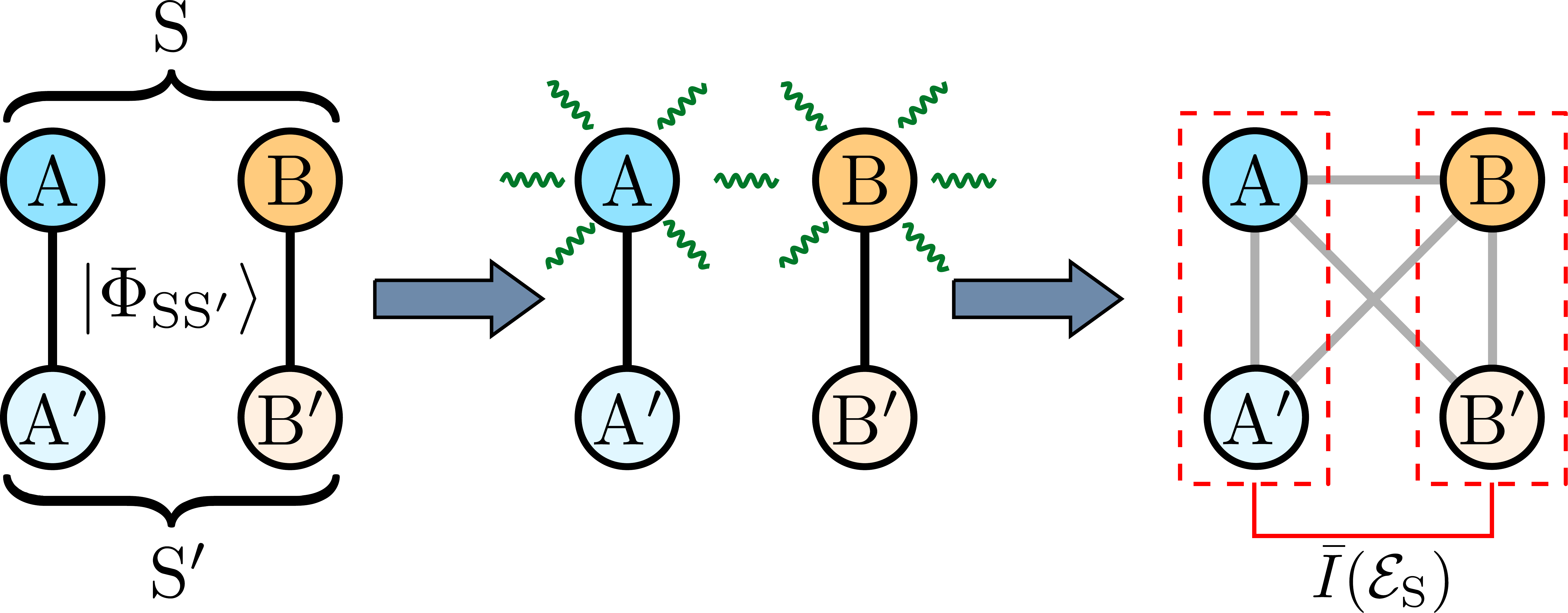}
\end{center}
\caption{Schematics of the method. Left: the system S is prepared in a maximally entangled state $\ket{\Phi_{\rmS\rmS'}}$ with the auxiliary system $\rmS'$ [this state is just a product of maximally entangled states between $\A\A'$ and $\B\B'$, see Eq. \eqref{MaxEnta}]. Middle: the system undergoes some dynamics $\mathcal{E}_\rmS$. Right: if and only if this process is correlated with respect to A and B, the total system $\rmS\rmS'$ becomes correlated with respect to the bipartition $\A\A'|\B\B'$ and the degree of correlation can be measured by the normalized mutual information, Eq. \eqref{Ibar}.}
\end{figure}
%%%%%%%%%%%%%%%%%%%%%%%%%%
%%%%%%%%%%%%%%%%%%%%%%%%%%

%%%%%%%%%%%%
\subsection{Construction of the correlation measure}
%%%%%%%%%%%%
In order to formulate a faithful measure of spatial correlations for dynamics, we adopt a resource theory approach \cite{MartinShash,Brandao1,Gour1,Brandao2,Emerson,Gour2,Tilmann,deVicente}. This is, we may consider correlated dynamics as a resource to perform whatever task that cannot be implemented solely by (composing) uncorrelated evolutions $\mathcal{E}_\A\otimes\mathcal{E}_\B$. Then, suppose that the system S undergoes some dynamics given by the map $\mathcal{E}_\rmS$, and consider the (left and right) composition of $\mathcal{E}_\rmS$ with some uncorrelated maps $\mathcal{L}_\A\otimes\mathcal{L}_\B$ and $\mathcal{R}_\A\otimes\mathcal{R}_\B$, so that the total dynamics is given by $\mathcal{E}'_\rmS=(\mathcal{L}_\A\otimes\mathcal{L}_\B)\mathcal{E}_\rmS(\mathcal{R}_\A\otimes\mathcal{R}_\B)$. It is clear that any task that we can do with $\mathcal{E}'_\rmS$ by composition with uncorrelated maps can also be achieved with $\mathcal{E}_\rmS$ by composition with uncorrelated maps. Hence, we assert that the amount of correlation in $\mathcal{E}_{\rm S}$ is at least as large as in $\mathcal{E}'_\rmS$. In other words, the amount of correlations of some dynamics does not increase under composition with uncorrelated dynamics. This is the fundamental law of this resource theory, and any faithful measure of correlations should satisfy it. For the sake of comparison, in the resource theory of entanglement, entanglement is the resource, and the fundamental law is that entanglement cannot increase under application of local operations and classical communication (LOCC) \cite{MartinShash}.

In this spirit, we introduce a measure of correlations for dynamics via the (normalized) quantum mutual information of the Choi-Jamio{\l}kowski state $\rho^{\rm CJ}_\rmS$, Eq. \eqref{CJstate},
\begin{align}\label{Ibar}
\bar{I}(\mathcal{E}_\rmS)&:=\frac{I(\rho^{\rm CJ}_\rmS)}{4 \log d}\\
&:=\frac{1}{4 \log d}\left[S\left(\rho^{\rm CJ}_\rmS|_{\A\A'}\right)+S\left(\rho^{\rm CJ}_\rmS|_{\B\B'}\right)-S\left(\rho^{\rm CJ}_\rmS\right)\right],\nonumber
\end{align}
with $S(\cdot):=-\Tr[(\cdot) \log(\cdot)]$ the von Neumann entropy evaluated for the reduced density operators $\rho^{\rm CJ}_\rmS|_{\A\A'}:=\Tr_{\B\B'}(\rho^{\rm CJ}_\rmS)$ and $\rho^{\rm CJ}_\rmS|_{\B\B'}:=\Tr_{\A\A'}(\rho^{\rm CJ}_\rmS)$, and $\rho^{\rm CJ}_\rmS$; see Fig. 1. The quantity $\bar{I}(\mathcal{E}_\rmS)$ is a faithful measure of how correlated the dynamics given by $\mathcal{E}_\rmS$ is, as it satisfies the following properties:

\begin{enumerate}
\item[i)] $\bar{I}(\mathcal{E}_\rmS)=0$ if and only if $\mathcal{E}_\rmS$ is uncorrelated, $\mathcal{E}_\rmS=\mathcal{E}_\A\otimes\mathcal{E}_\B$. This follows from the fact that the Choi-Jamio{\l}kowski state of an uncorrelated map is a product state with respect to the bipartition $\A\A'|\B\B'$, see Appendix \ref{Ap:1}.
\item[ii)] $\bar{I}(\mathcal{E}_\rmS)\in[0,1]$. It is clear that $\bar{I}(\mathcal{E}_\rmS)\geq0$, moreover it reaches its maximum value when $S(\rho^{\rm CJ}_\rmS)$ is minimal and $S\left(\rho^{\rm CJ}_\rmS|_{\A\A'}\right)+S\left(\rho^{\rm CJ}_\rmS|_{\B\B'}\right)$ is maximal. Both conditions meet when $\rho^{\rm CJ}_\rmS$ is a maximally entangled state with respect to the bipartition $\A\A'|\B\B'$, leading to $I(\rho^{\rm CJ}_\rmS)=2\log d^2$.
\item[iii)] The fundamental law is satisfied,
\begin{equation}\label{fundamentallaw}
\bar{I}(\mathcal{E}_\rmS)\geq\bar{I}[(\mathcal{L}_\A\otimes\mathcal{L}_\B) \mathcal{E}_\rmS (\mathcal{R}_\A\otimes\mathcal{R}_\B)],
\end{equation}
where the equality is reached for uncorrelated unitaries $\mathcal{L}_{\A}(\cdot)=U_{\A}(\cdot)U^\dagger_{\A}$, $\mathcal{L}_{\B}(\cdot)=U_{\B}(\cdot)U^\dagger_{\B}$, $\mathcal{R}_{\A}(\cdot)=V_{\A}(\cdot)V^\dagger_{\A}$, and $\mathcal{R}_{\B}(\cdot)=V_{\B}(\cdot)V^\dagger_{\B}$. This result follows from the monotonicity of the quantum mutual information under local CPT maps (which in turn follows from the monotonicity of quantum relative entropy \cite{Vedral}) and the fact that for any matrix $A$, $A\otimes\mathbb{1}_{\rmS'}\ket{\Phi_{\rmS\rmS'}}=\mathbb{1}_{\rmS}\otimes A^{\rm t}\ket{\Phi_{\rmS\rmS'}}$ where the superscript ``t'' denotes the transposition in the Schmidt basis of the maximally entangled state $\ket{\Phi_{\rmS\rmS'}}$.
\end{enumerate}

%%%%%%%%%%%%
\subsection{Maximally correlated dynamics}\label{sec:MCD}
%%%%%%%%%%%%
Before computing $\bar{I}$ for some cases it is worth studying which dynamics achieve the maximum value $\bar{I}_{\rm max}=1$. From the resource theory point of view, these dynamics can be considered as maximally correlated since they cannot be constructed from other maps by composition with uncorrelated maps [because of Eq. \eqref{fundamentallaw}]. We have the following results:

\smallskip
\noindent \textit{Theorem 1}. If for a map $\mathcal{E}_\rmS$ the property $\bar{I}(\mathcal{E}_\rmS)=1$ holds, such map must be unitary $\mathcal{E}_\rmS(\cdot)=U_\rmS(\cdot)U_\rmS^\dagger$, $U_\rmS U_\rmS^\dagger=\mathbb{1}$.

\smallskip
\noindent \textit{Proof}. As aforementioned, the maximum value, $\bar{I}(\mathcal{E}_\rmS)=1$, is reached if and only if $\rho^{\rm CJ}_{\rmS}$ is a maximally entangled state with respect to the bipartition $\A\A'|\B\B'$, $|\Psi_{(\A\A')|(\B\B')}\rangle$. Then
\begin{equation}
\mathcal{E}_\rmS\otimes\mathds{1}_{\rmS'}(\ket{\Phi_{\rmS\rmS'}}\bra{\Phi_{\rmS\rmS'}})=|\Psi_{(\A\A')|(\B\B')}\rangle\langle\Psi_{(\A\A')|(\B\B')}|
\end{equation}
is a pure state. Therefore $\mathcal{E}_\rmS$ must be unitary as the Choi-Jamio{\l}kowski state is pure if and only if it represents a unitary map. \qed

Despite the connection with maximally entangled states, the set of maximally correlated operations $\mathfrak{C}:=\{U_\rmS; \bar{I}(U_\rmS)=1\}$, can not be so straightforwardly characterized as it may seem. Note that not all maximally entangled states $|\Psi_{(\A\A')|(\B\B')}\rangle$ are valid Choi-Jamio{\l{kowski states. In Appendix \ref{Ap:2} we provide a detailed proof of the next theorem.

\smallskip
\noindent \textit{Theorem 2}. A unitary map $U_\rmS\in\mathfrak{C}$ if and only if it fulfills the equation
\begin{equation}\label{Umax3}
\sum_{i,j} \langle k i |U_{\rmS}| m j  \rangle\langle n j  |U_{\rmS}^\dagger| \ell i \rangle=\delta_{k\ell}\delta_{mn}.
\end{equation}

Examples of maximally correlated dynamics are the swap operation exchanging the states of the two parties A and B, $U_{\rmS}=U_{\A\leftrightarrow\B}$, and thus also any unitary of the form of $(U_{\A}\otimes U_{\B}) U_{\A\leftrightarrow\B}(V_{\A}\otimes V_{\B})$. However, not every $U_\rmS\in\mathfrak{C}$ falls into this class. For example, the unitary operation of two qubits $U'_{\rmS}=|21\rangle\langle12|+{\rm i}(|11\rangle\langle21|+|12\rangle\langle11|+|22\rangle\langle22|)$ belongs to $\mathfrak{C}$ and it cannot be written as $(U_{\A}\otimes U_{\B}) U_{\A\leftrightarrow\B}(V_{\A}\otimes V_{\B})$, since that would imply vanishing $\bar{I}(U'_{\rmS}U_{\A\leftrightarrow\B})$ whereas $\bar{I}(U'_{\rmS}U_{\A\leftrightarrow\B})=1/2\neq0$.
Interestingly, operations able to create highly correlated states such as the two-qubit controlled-NOT gate \cite{NC00} as well as the two-qubit dynamical maps describing the dissipative generation of Bell states \cite{Barreiro, Mueller} achieve a correlation value of 1/2 and thus do not correspond to maximally correlated dynamics. Note that whereas a controlled-NOT gate creates for appropriately chosen two-qubit initial states maximally entangled states, there are other states which are left completely uncorrelated under its action. The measure $\bar{I}$ captures - completely independently of initial states and of whether possibly created correlations are quantum or classical - the fact that correlated dynamics cannot be realized by purely local dynamics.

Let us point out that in some resource theories, such as bi-partite entanglement, the maximal element can generate any other element by applying the operations which fulfill its fundamental law, e.g. LOCC. This is not the case here, i.e. maximally correlated evolutions cannot generate any arbitrary dynamics by composition with uncorrelated operations. Indeed, if $\mathcal{E}_\rmS^{\rm max}$ were able to generate any other dynamics, in particular it would be able to generate any unitary evolution $U_\rmS$,  $(\mathcal{L}_\A\otimes\mathcal{L}_\B)\mathcal{E}_\rmS^{\rm max}(\mathcal{R}_\A\otimes\mathcal{R}_\B)(\cdot)=U_\rmS(\cdot)U_\rmS^\dagger$. However, this would imply that $\mathcal{L}_\A\otimes\mathcal{L}_\B$, $\mathcal{E}_\rmS^{\rm max}$ and $(\mathcal{R}_\A\otimes\mathcal{R}_\B)$ are unitary evolutions as well, so that $(U_\A\otimes U_\B ) U^{\rm max}_\rmS (V_\A\otimes V_\B )=U_\rmS$, with $\mathcal{E}_\rmS^{\rm max}(\cdot)=U^{\rm max}_\rmS(\cdot)U_\rmS^{\rm max\dagger}$. Since $\bar{I}(\mathcal{E}_\rmS)$ is invariant under the composition of uncorrelated unitaries, this result would imply that for any correlated unitary $U_\rmS$, $\bar{I}(U_\rmS)$ would take the same value $[\bar{I}(U^{\rm max}_\rmS)]$, and this is not true as can be easily checked.

%%%%%%%%%%%%%%%%%%%%%%%%%%
%%%%%%%%%%%%%%%%%%%%%%%%%%
\begin{figure}[t]
	\includegraphics[width=\columnwidth]{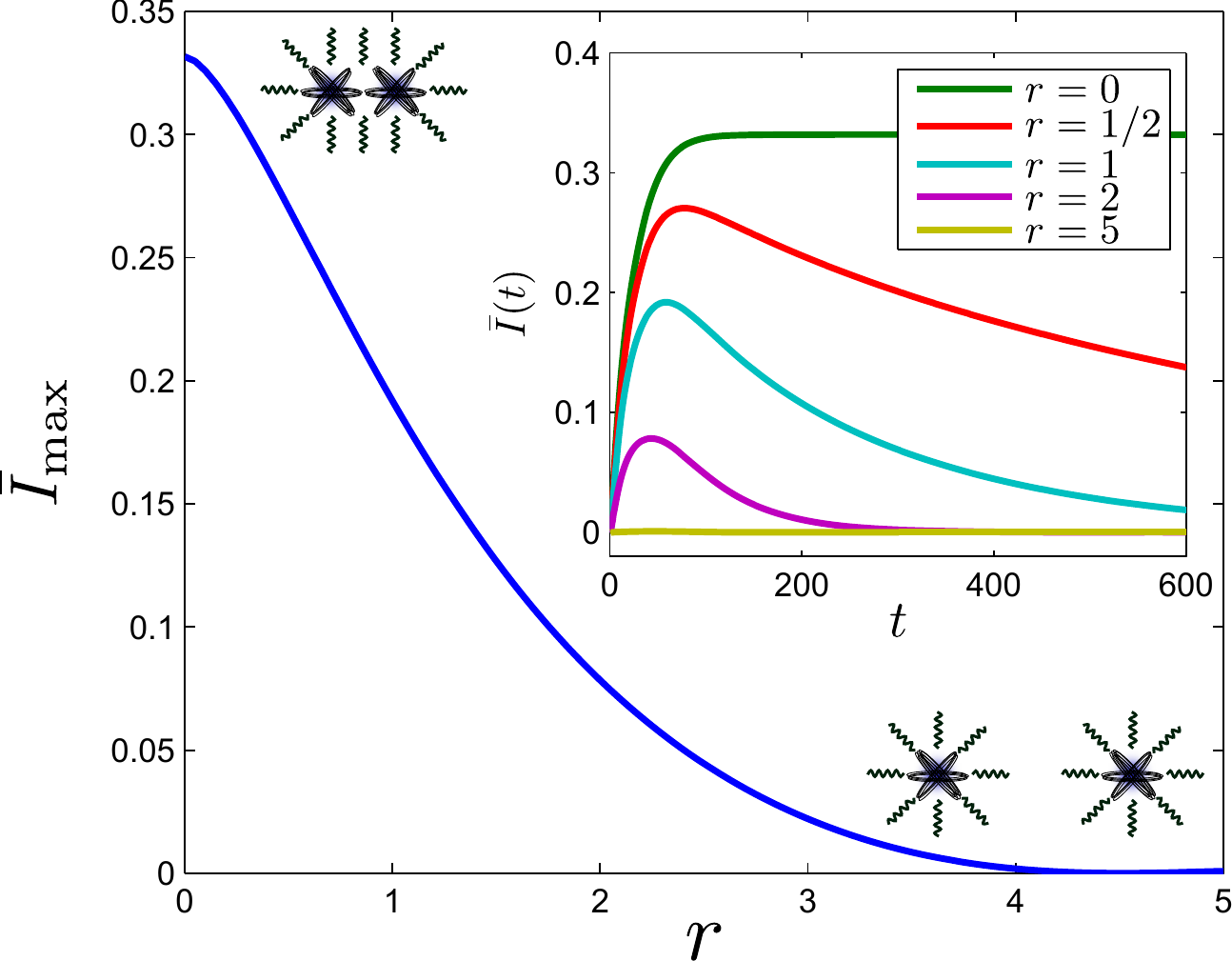}
	\caption{Maximum value of $\bar{I}$ as a function of the distance $r$ for two two-level atoms radiating in the electromagnetic vacuum. As expected, the amount of correlations in the dynamics decreases with $r$. In the inset, $\bar{I}$ is represented as a function of time for different distances $r$ between atoms ($\omega=|\bm{d}|/2=1$, $\theta=0$, see Appendix \ref{Ap:3}).}
\label{fig2}
\end{figure}
%%%%%%%%%%%%%%%%%%%%%%%%%%
%%%%%%%%%%%%%%%%%%%%%%%%%%

%%%%%%%%%%%%
\section{Applications}
%%%%%%%%%%%%

%%%%%%%%%%%%
\subsection{Two-level atoms in the electromagnetic vacuum}
%%%%%%%%%%%%
To illustrate the behavior of $\bar{I}(\mathcal{E}_\rmS)$, consider the paradigmatic example of two identical two-level atoms with transition frequency $\omega$ interacting with the vacuum of the electromagnetic radiation field (see Appendix \ref{Ap:3}). Under a series of standard approximations, the dynamics of the reduced density matrix of the atoms $\rho_\rmS$ is described by the master equation
%%%
\begin{align}
\frac{d\rho_\rmS}{dt}=\mathcal{L}(\rho_\rmS)=&-\ii\tfrac{\omega}{2}[\sigma_1^z+\sigma_2^z,\rho_\rmS]\\
&+\sum_{j,k=1,2}a_{jk}\Big(\sigma_k^-\rho_{\rmS}\sigma_j^+-\tfrac{1}{2}\{\sigma_j^+\sigma_k^-,\rho_\rmS\}\Big),\nonumber
\end{align}
%%%
where $\sigma^z_j$ is the Pauli $z$-matrix for the $j$-th atom, and $\sigma_j^+=(\sigma_j^-)^\dagger=|e\rangle\mbox{}_j\langle g|$ the electronic raising and lowering operators, describing transitions between the exited $\ket{e}_j$ and ground $\ket{g}_j$ states. The coefficients $a_{jk}$ depend on the spatial separation $r$ between the atoms. In the limit of $r\gg 1/\omega$ they reduce to $a_{jk}\simeq \gamma_0\delta_{jk}$, whereas for $r\ll 1/\omega$ they take the form $a_{jk}\simeq\gamma_0$. Here $\gamma_0$ is the decay rate of the individual transition between $\ket{e}$ and $\ket{g}$.
In the first regime the two-level atoms interact effectively with independent environments, while in the second, the transitions are collective and lead to the Dicke model of super-radiance \cite{Dicke}.

To quantitatively assess this behavior of uncorrelated/correlated dynamics as a function of $r$, we compute the measure of correlations $\bar{I}$, Eq. \eqref{Ibar} (see Appendix \ref{Ap:3} for details). The results are shown in Fig. 2. Despite the fact that the value of $\bar{I}$ depends on time (the dynamical map is $\mathcal{E}_{\rm S}={\rm e}^{t\mathcal{L}}$), $\bar{I}$ decreases as $r$ increases, as expected. Furthermore, the value of $\bar{I}$ approaches zero for $t$ large enough (see inset plot), except in the limiting case $r=0$, because for $r\neq0$ the dynamics becomes uncorrelated in the asymptotic limit, $\lim_{t\rightarrow\infty}{\rm e}^{t\mathcal{L}}=\mathcal{E}\otimes\mathcal{E}$, where $\mathcal{E}(\cdot)=K_1(\cdot)K_1^\dagger+K_2(\cdot)K_2^\dagger$ with Kraus operators $K_1=\bigl( \begin{smallmatrix}
  0 & 0\\
  1 & 0
\end{smallmatrix} \bigr)$ and $K_2=\bigl( \begin{smallmatrix}
  0 & 0\\
  0 & 1
\end{smallmatrix} \bigr)$; however for $r=0$, $\lim_{t\rightarrow\infty}{\rm e}^{t\mathcal{L}}$ is a correlated map. Thus, we obtain perfect agreement between the rigorous measure of correlations $\bar{I}$ and the physically expected behavior of two distant atoms undergoing independent noise.

%%%%%%%%%%%%
\subsection{Spatial noise correlations in quantum computing}
%%%%%%%%%%%%
Fault-tolerant quantum computing is predicted to be achievable provided that detrimental noise is sufficiently weak \textit{and} not too strongly correlated \cite{Preskill}. However, even if noise correlations decay sufficiently fast in space, associated (provable) bounds for the accuracy threshold values can decrease by several orders of magnitude as compared to uncorrelated noise \cite{Kitaev}. Thus, it is of both fundamental and practical importance \cite{Joe} to be able to detect, quantify and possibly reduce without a priori knowledge of the underlying microscopic dynamics the amount of correlated noise. Here, we exemplify how the proposed measure can be employed in this context by applying it to a simple, though paradigmatic model system of two representative qubits from a larger qubit register. We assume that the qubits are exposed to local thermal (bosonic) baths, such as realized e.g.~by coupling distant atomic qubits to the surrounding electromagnetic radiation field, and that they interact via a weak ZZ-coupling, which could be caused, e.g., by undesired residual dipolar or van-der-Waals type interactions between the atoms. The ``error'' dynamics of this system is described by the master equation
\begin{align}\label{masterZZ}
\frac{d\rho_\rmS}{dt}=\mathcal{L}(\rho_\rmS)=&-\ii[\tfrac{\omega}{2}(\sigma_1^z+\sigma_2^z)+J\sigma^z_1\sigma^z_2,\rho_\rmS]\\
+&\sum_{j=1,2}\gamma_0(\bar{n}+1)\Big(\sigma_j^-\rho_{\rmS}\sigma_j^+-\tfrac{1}{2}\{\sigma_j^+\sigma_j^-,\rho_\rmS\}\Big)\nonumber\\
+&\sum_{j=1,2}\gamma_0\bar{n}\Big(\sigma_j^+\rho_{\rmS}\sigma_j^--\tfrac{1}{2}\{\sigma_j^-\sigma_j^+,\rho_\rmS\}\Big),\nonumber
\end{align}
where $\omega$ is the energy difference between the qubit states, $J$ the strength of the residual Hamiltonian coupling, $\gamma_0$ is again the decay rate between upper and lower energy level of each individual qubit and $\bar{n}=[\exp(\omega/T)-1]^{-1}$ is mean number of bosons with frequency $\omega$ in the two local baths of temperature $T$ (assumed to be equal).

We assume $J$ and $\gamma_0$ to be out of our control and aim at studying the spatial correlations of the errors induced by the interplay of the residual ZZ-coupling and the baths as a function of the bath temperature $T$ and elapsed time $t$, which in the present context might be interpreted as the time for executing one round of quantum error correction \cite{Preskill,Dennis}. Since the overall probability that some error occurs on the two qubits will increase under increasing $t$ and $T$, we need to fix it for a fair assessment of the correlation of the dynamics. A natural way to do this is by defining the error probability in terms of how close the dynamical map induced by Eq.~\eqref{masterZZ} [excluding the term $\tfrac{\omega}{2}(\sigma_1^z+\sigma_2^z)$, as this is not considered a source of error] is to the identity map (the case of no errors). Particularly, we can use the fidelity between both Choi-Jamio{\l}kowski states, $\rho^{\rm CJ}_{\rm S}$ for the ``error'' map and $|\Phi_{\rmS\rmS'}\rangle$ for the identity map,
$P_{\rm error}=1-\sqrt{\langle\Phi_{\rmS\rmS'}|\rho^{\rm CJ}_{\rm S}|\Phi_{\rmS\rmS'}\rangle}$. Figure 3 shows the value of amount of dynamical correlations as measured by $\bar{I}$ along a $t$-$T$ line on which the error probability is constant ($P_{\rm error}=0.1$, green line in the inset plot). The numerical data shows, despite this fixing of the overall error rate, that as the temperature increases the correlatedness of errors decreases very rapidly. This remarkable result suggests that by increasing the effective, surrounding temperature one can strongly decrease the non-local character of the noise at the expense of a slightly higher error rate per fixed time $t$, or constant error rates if the time $t$ for an error correction round can be reduced. Thus, the proposed quantifier might prove useful to meet and certify in a given physical architecture the noise levels and noise correlation characteristics which are required to reach the regime where fault-tolerant scalable quantum computing becomes feasible in practice.

%%%%%%%%%%%%%%%%%%%%%%%%%%
%%%%%%%%%%%%%%%%%%%%%%%%%%
\begin{figure}[t]
\begin{center}
\includegraphics[width=\columnwidth]{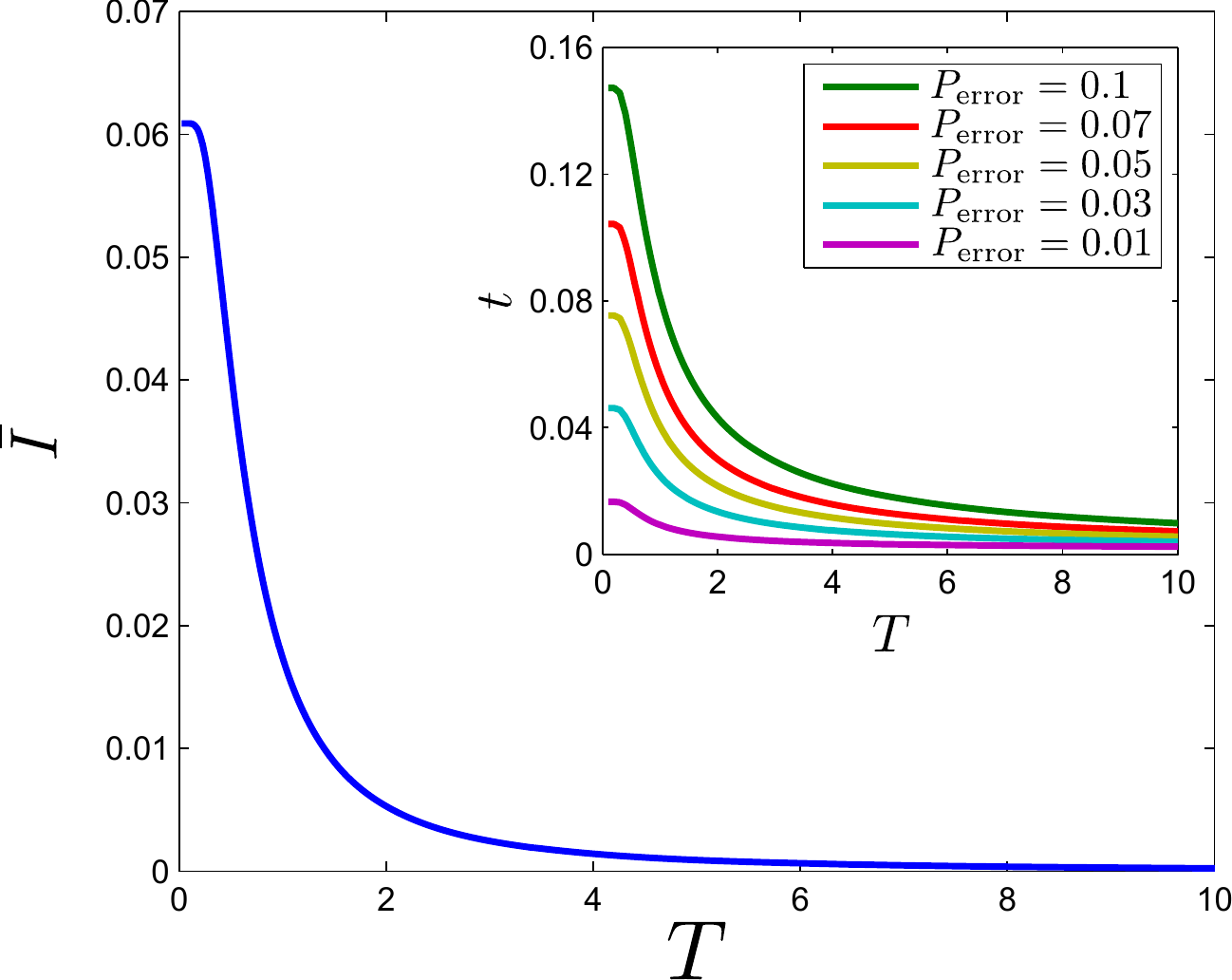}
\end{center}
\caption{Amount of spatial correlations $\bar{I}$ along the $t$--$T$ line corresponding to constant error probability $P_{\rm error}=0.1$. We see the rapid decreasing of $\bar{I}$ as $T$ increases ($J=1$ and $\gamma_0=4/3$ in units of $\omega$). The inset shows $t$--$T$ isolines for various values of the error probability $P_{\rm error}$, which increases with both $t$ and $T$.}
\end{figure}
%%%%%%%%%%%%%%%%%%%%%%%%%%
%%%%%%%%%%%%%%%%%%%%%%%%%%

%%%%%%%%%%%%
\section{Conclusion}
%%%%%%%%%%%%

In this work, we have formulated a general measure for the spatial correlations of quantum dynamics without restriction to any specific model. To that aim we have adopted a resource theory approach and obtained a fundamental law that any faithful quantifier of spatial correlation must satisfy. We have characterized the maximally correlated dynamics, and applied our measure to the paradigmatic example of two atoms radiating in the electromagnetic field, where spatial correlations are naturally related to the separation between atoms. Furthermore, we have illustrated the applicability of the measure in the context of quantum computing, where it can be employed to quantify and potentially control spatial noise correlations without a priori knowledge of the underlying dynamics.

Beyond the scope of this work it will be interesting from a fundamental point of view to study how many independent (up to local unitaries) maximally correlated dynamics there are, and how to deal with the case of multi-partite or infinite dimensional systems. From a practical point of view, it is also interesting to develop efficient methods to estimate the proposed measure, in particular in high-dimensional quantum systems, e.g.~by the construction of witnesses or bounds, in analogy to entanglement estimators \cite{GuhneToth} that have been developed based on the resource theory of entanglement. In this regard, it is our hope that the present results provide a useful tool to study rigorously the role of spatial correlations in a variety of physical processes, including noise assisted transport, quantum computing and dissipative phase transitions.

\section*{Acknowledgments}

We acknowledge interesting discussions with T. Monz and D. Nigg, as well as financial support by the Spanish MINECO grant FIS2012-33152, the CAM research consortium QUITEMAD grant S2009-ESP-1594, the European Commission PICC: FP7 2007-2013, Grant No.~249958, the UCM-BS grant GICC-910758 and the U.S. Army Research Office through grant W911NF-14-1-0103.

\appendix

\section{Choi-Jamio{\l}kowski state of uncorrelated maps}
\label{Ap:1}
First of all, let $U_{\B\leftrightarrow\A'}$ be the commutation matrix (or unitary swap operation) \cite{vec1,vec2} between Hilbert subspaces $\calH_\B$ and $\calH_{\A'}$ of the total Hilbert space $\calH_\A\otimes \calH_\B\otimes \calH_{\A'}\otimes \calH_{\B'}$:
\begin{eqnarray}
U_{\B\leftrightarrow\A'} \left(M_1\otimes M_2\otimes M_{3}\otimes M_{4}\right) U_{\B\leftrightarrow\A'}^\dagger\\
= M_1\otimes M_{3} \otimes M_2\otimes M_{4}.
\end{eqnarray}
where $M_1$, $M_2$, $M_3$ and $M_4$ are operators acting on the respective Hilbert subspaces in the decomposition $\calH_\A\otimes \calH_\B\otimes \calH_{\A'}\otimes \calH_{\B'}$. This is, $M_1$ acts on $\calH_\A$, $M_4$ on $\calH_{\B'}$, and $M_2$ and $M_{3}$ act on $\calH_\B$ and $\calH_{\A'}$ on the left hand side and on $\calH_{\A'}$ and $\calH_\B$ on the right hand side of the equality respectively. Note that $U_{\B\leftrightarrow\A'}U_{\B\leftrightarrow\A'}=\mathbb{1}$ and then $U_{\B\leftrightarrow\A'}=U_{\B\leftrightarrow\A'}^\dagger$.

Now, it turns out that the evolution given by some dynamical map $\mathcal{E}_\rmS$ is uncorrelated with respect to the subsystems $\A$ and $\B$, $\mathcal{E}_\rmS=\mathcal{E}_\A\otimes \mathcal{E}_\B$, if and only if its Choi-Jamio{\l}kowski state $\rho^{\rm CJ}_\rmS:=\mathcal{E}_\rmS\otimes\mathds{1}_{\rmS'}(\ket{\Phi_{\rmS\rmS'}}\bra{\Phi_{\rmS\rmS'}})$ is
\begin{equation}\label{CJstatetilde}
\rho^{\rm CJ}_\rmS=U_{\B\leftrightarrow\A'}\left(\rho^{\rm CJ}_\A\otimes \rho^{\rm CJ}_\B\right) U_{\B\leftrightarrow\A'},
\end{equation}
where $\rho^{\rm CJ}_\A$ and $\rho^{\rm CJ}_\B$ are the Choi-Jamio{\l}kowski states of the maps $\mathcal{E}_\A$ and $\mathcal{E}_\B$, respectively.

Indeed, if $\mathcal{E}_\rmS=\mathcal{E}_\A\otimes \mathcal{E}_\B$, we have (omitting for the sake of clarity the subindexes in the basis expansion of $\ket{\Phi_{\rmS\rmS'}}$):
\begin{widetext}
\begin{align}
\rho^{\rm CJ}_\rmS=\mathcal{E}_\rmS\otimes\mathds{1}_{\rmS'}(\ket{\Phi_{\rmS\rmS'}}\bra{\Phi_{\rmS\rmS'}})&=\frac{1}{d^2}\sum_{k,\ell,m,n=1}^d\mathcal{E}_\rmS\left(|k\ell\rangle\langle mn|\right)\otimes|k\ell\rangle\langle mn| \nonumber\\
&=\frac{1}{d^2}\sum_{k,\ell,m,n=1}^d\mathcal{E}_\A\left(|k\rangle\langle m|\right)\otimes\mathcal{E}_\B\left(|\ell\rangle\langle n|\right)\otimes|k\ell\rangle\langle mn|,
\end{align}
then
\begin{align}
U_{\B\leftrightarrow\A'}\rho^{\rm CJ}_\rmS U_{\B\leftrightarrow\A'}&=\frac{1}{d^2}\sum_{k,\ell,m,n=1}^d\mathcal{E}_\A(|k\rangle\langle m|)\otimes|k\rangle\langle m|\otimes\mathcal{E}_\B(|\ell\rangle\langle n|)\otimes |\ell\rangle\langle n|\nonumber\\
&=\frac{1}{d}\sum_{k,m=1}^d\mathcal{E}_\A\otimes\mathds{1}(|kk\rangle\langle mm|)\otimes\frac{1}{d}\sum_{\ell,n=1}^d\mathcal{E}_\B\otimes\mathds{1}(|\ell \ell\rangle\langle nn|)\nonumber\\
&=\rho^{\rm CJ}_\A\otimes \rho^{\rm CJ}_\B.
\end{align}
\end{widetext}
Conversely, if Eq. \eqref{CJstatetilde} holds, then the dynamics has to be uncorrelated because the correspondence between Choi-Jamio{\l}kowski states and dynamical maps is one-to-one.

From Eq. \eqref{CJstatetilde} it is straightforward to conclude that $\bar{I}(\mathcal{E}_{\rmS})=0$ if and only if $\mathcal{E}_{\rmS}$ is uncorrelated, because the von Neumann entropy of the Choi-Jami{\l}kowski state factorizes $S(\rho^{\rm CJ}_\rmS)=S\left[U_{\B\leftrightarrow\A'}\left(\rho^{\rm CJ}_\A\otimes \rho^{\rm CJ}_\B\right)U_{\B\leftrightarrow\A'}\right]=S\left(\rho^{\rm CJ}_\A\otimes \rho^{\rm CJ}_\B\right)=S\left(\rho^{\rm CJ}_\A\right)+S\left(\rho^{\rm CJ}_\B\right)$ if and only if $\mathcal{E}_{\rmS}$ is uncorrelated.

\section{Proof of Theorem 2} \label{Ap:2}
As commented in section \ref{sec:MCD}, $U_\rmS\in\mathfrak{C}$ if
\begin{equation}\label{proof1}
\ket{\Psi_{(\A\A')|(\B\B')}}=U_{\rmS}\otimes\mathbb{1}|\Phi_{\rmS\rmS'}\rangle,
\end{equation}
where $\ket{\Psi_{(\A\A')|(\B\B')}}$ is a maximally entangled state with respect to the bipartition $\A\A'|\B\B'$. Note that if $\ket{\Psi_{(\A\A')|(\B\B')}}$ is a maximally entangled state with respect to the bipartition $\A\A'|\B\B'$, $U_{\B\leftrightarrow\A'}\ket{\Psi_{(\A\A')|(\B\B')}}$ will be a maximally entangled state state with respect to the bipartition $\A\B|\A'\B'=\rmS|\rmS'$. Since any maximally entangled state with respect to the bipartition $\rmS|\rmS'$ can be written as $\tilde{U}_{\rmS}\otimes\tilde{U}_{\rmS'}|\Phi_{\rmS\rmS'}\rangle$ for some local unitaries $\tilde{U}_{\rmS}$ and $\tilde{U}_{\rmS'}$, we can write
\begin{equation}\label{proof2}
U_{\B\leftrightarrow\A'}\ket{\Psi_{(\A\A')|(\B\B')}}=\tilde{U}_{\rmS}\otimes\tilde{U}_{\rmS'}|\Phi_{\rmS\rmS'}\rangle.
\end{equation}
Because of Eqs. \eqref{proof1} and \eqref{proof2} we conclude that $U_{\rmS}\in\mathfrak{C}$ if and only if there exist unitaries $\tilde{U}_{\rmS}$ and $\tilde{U}_{\rmS'}$ such that
\begin{equation}\label{Umax1}
U_{\rmS}\otimes\mathbb{1}_{\rmS'}|\Phi_{\rmS\rmS'}\rangle=U_{\B\leftrightarrow\A'} ( \tilde{U}_{\rmS}\otimes\tilde{U}_{\rmS'} ) |\Phi_{\rmS\rmS'}\rangle.
\end{equation}
Next, we prove the following

\smallskip

\noindent \textit{Lemma}. A unitary map $U_\rmS\in\mathfrak{C}$ if and only if there exists some other unitary $V$ such that the matrix elements of $U_S$ can be written as
\begin{equation}\label{Umax2}
\langle k\ell|U_{\rmS}|m n\rangle=\langle k m |V|\ell n \rangle.
\end{equation}
\noindent \textit{Proof}. If $U_\rmS\in\mathfrak{C}$, then by taking inner product with respect to the basis element $\ket{ k \ell m n}$ in Eq. \eqref{Umax1} we obtain:
\begin{align}\label{Vintermedio}
\langle k\ell|U_{\rmS}|m n\rangle&=d\langle k m \ell n |\tilde{U}_{\rmS}\otimes \tilde{U}_{\rmS'}|\Phi_{\rmS\rmS'}\rangle \nonumber\\
&=\langle k m |\tilde{U}_{\rmS} \tilde{U}^{\rm t}_{\rmS'}|\ell n \rangle=\langle k m |V|\ell n\rangle,
\end{align}
for $V=\tilde{U}_{\rmS} \tilde{U}^{\rm t}_{\rmS'}$. Here we have used that $A\otimes\mathbb{1}_{\rmS'}\ket{\Phi_{\rmS\rmS'}}=\mathbb{1}_{\rmS}\otimes A^{\rm t}\ket{\Phi_{\rmS\rmS'}}$ where the superscript ``t'' denotes the transposition in the Schmidt basis of the maximally entangled state $\ket{\Phi_{\rmS\rmS'}}$, which has been taken to be the canonical basis here.

\noindent Conversely, assume that there exists a unitary $V$ satisfying \eqref{Umax2}. As $V$ can always be decomposed as the product of two unitaries, %\cite{decompositionV}
 $V=V_1V_2$, by setting $\tilde{U}_{\rmS}=V_1$ and $\tilde{U}_{\rmS'}^{\rm t}=V_2$, the same algebra as in Eq. \eqref{Vintermedio} leads us to rewrite Eq. \eqref{Umax2} as
\begin{equation}
\langle k \ell m n|U_{\rmS}\otimes \mathbb{1}_{\rmS'}|\Phi_{\rmS\rmS'}\rangle=\langle k \ell m n|U_{\B\leftrightarrow\A'}\tilde{U}_{\rmS}\otimes \tilde{U}_{\rmS'}|\Phi_{\rmS\rmS'}\rangle.
\end{equation}
Since $\ket{ k \ell m n}$ are elements of a basis we conclude that Eq. \eqref{Umax1} holds. \qed

\smallskip

\noindent With these results, the Theorem 2 is easy to prove.

\smallskip

\noindent \textit{Proof of Theorem 2}. Note that for any unitary $U_\rmS$, Eq. \eqref{Umax2} is satisfied for some matrix $V$. Thus, what we have to prove is that such a matrix $V$ is unitary if and only if $U_\rmS$ fulfills the equation
\begin{equation}
\sum_{i,j} \langle k i |U_{\rmS}| m j  \rangle\langle n j  |U_{\rmS}^\dagger| \ell i \rangle=\delta_{k\ell}\delta_{mn},
\end{equation}
 and this follows after a straightforward algebraic computation. \qed

\section{Two two-level atoms coupled to the radiation field}
\label{Ap:3}

The free Hamiltonian of the atoms is
\begin{equation}
H_\rmS=\frac{\omega}{2}(\sigma^z_1+\sigma^z_2),
\end{equation}
where $\sigma^z_j$ is the Pauli $z$-matrix for the $j$-th atom. In addition, the environmental free Hamiltonian is given by
\begin{equation}
H_{\rm E}=\sum_{\bm{k}} \sum_{\lambda =1,2} \omega_{\bm{k}} a_\lambda^\dagger(\bm{k})a_\lambda(\bm{k}),
\end{equation}
where $\bm{k}$ and $\lambda$ stand for the wave vector and the two polarization degrees of freedom, respectively. We have taken natural units $\hbar=c=1$. The dispersion relation in the free space is $\omega_{\rm{k}}=|\bm{k}|$, and the field operators $a_\lambda^\dagger(\bm{k})$ and $a_\lambda(\bm{k})$ describe the creation and annihilation of photons with wave vector $\bm{k}$ and polarization vector $\bm{e}_\lambda$. These fulfill
$\bm{k}\cdot\bm{e}_\lambda=0$ and $\bm{e}_\lambda\cdot\bm{e}_{\lambda'}=\delta_{\lambda,\lambda'}$.

The atom-field interaction is described in dipole approximation by the Hamiltonian
\begin{equation}
H_{\rm SE}=-\sum_{j=1,2}\left[\sigma_j^-\bm{d}\cdot\bm{E}(\bm{r}_j)+\sigma_j^+\bm{d}^\ast\cdot\bm{E}(\bm{r}_j)\right].
\end{equation}
Here, $\bm{d}$ is the dipole matrix element of the atomic transition, $\bm{r}_j$ denotes the position of the $j$-th atom, and $\sigma_j^+=(\sigma_j^-)^\dagger=|e\rangle\mbox{}_j\langle g|$ for its exited $\ket{e}_j$ and ground $\ket{g}_j$ states. Furthermore, the electric field operator is given by (Gaussian units)
\begin{equation}
\bm{E}(\bm{r})=\ii\sum_{\bm{k},\lambda}\sqrt{\frac{2\pi\omega_{\bm{k}}}{\mathcal{V}}}\bm{e}_\lambda(\bm{k})\left(a_\lambda(\bm{k})\ee^{\ii \bm{k}\cdot\bm{r}}-a^\dagger_\lambda(\bm{k})\ee^{-\ii \bm{k}\cdot\bm{r}}\right),
\end{equation}
where $\mathcal{V}$ denotes the quantization volume.
In the Markovian weak coupling limit \cite{BrPe02} the master equation for the atoms takes the form:
\begin{align}\label{MasterApp}
\frac{d\rho_\rmS}{dt}=\mathcal{L}(\rho_\rmS)=&-\ii\tfrac{\omega}{2}[\sigma_1^z+\sigma_2^z,\rho_\rmS]\\
&+\sum_{i,j=1,2}a_{jk}\Big(\sigma_k^-\rho_{\rmS}\sigma_j^+-\tfrac{1}{2}\{\sigma_j^+\sigma_k^-,\rho_\rmS\}\Big),\nonumber
\end{align}
where, after taking the continuum limit ($\tfrac{1}{\mathcal{V}}\sum_{\bm{k}}\rightarrow \tfrac{1}{(2\pi)^3}\int d^3\bm{k}$) and performing the integrals, the coefficients $a_{jk}$ are given by (sec. 3.7.5 of \cite{BrPe02})
\begin{equation}
a_{jk}=\gamma_0[j_0(x_{jk})+P_2(\cos\theta_{jk})j_2(x_{jk})],
\end{equation}
here $\gamma_0=\frac{4}{3}\omega^3|\bm{d}|^2$, and $j_0(x)$ and $j_2(x)$ are spherical Bessel functions \cite{AS},
\begin{equation}
j_0(x)=\frac{\sin x}{x}, \quad j_2(x)=\left(\frac{3}{x^3}-\frac{1}{x}\right)\sin x-\frac{3}{x^2}\cos x,
\end{equation}
and
\begin{equation}
P_2(\cos\theta)=\frac{1}{2}(3\cos^2\theta- 1)
\end{equation}
is a Legendre polynomial, with
\begin{equation}
x_{jk}=\omega|\bm{r}_j-\bm{r}_k|, \quad \text{and } \cos^2(\theta_{jk})=\frac{|\bm{d}\cdot(\bm{r}_j-\bm{r}_k)|^2}{|\bm{d}|^2|\bm{r}_j-\bm{r}_k|^2}.
\end{equation}

Notice that if the distance between atoms $r=|\bm{r}_1-\bm{r}_2|$, is much larger than the wavelength associated with the atomic transition $r\gg 1/\omega$, we have $a_{jk}\simeq \gamma_0\delta_{ij}$ and only the diagonal terms $\gamma_0=\frac{4}{3}\omega^3|\bm{d}|^2$ are relevant. Then, the master equation describes two-level atoms interacting with independent environments, and there are no correlations in the emission of photons by the first and the second atom. In the opposite case, when  $r\ll 1/\omega$, every matrix element approaches the same value $a_{ij}\simeq\gamma_0$, in the master equation the atomic transitions can be approximately described by the collective jump operators $J_{\pm}=\sigma_1^\pm+\sigma_1^\pm$, and the pair of atoms becomes equivalent to a four-level system with Hamiltonian $\omega J_z=\frac{\omega}{2} (\sigma_1^z+\sigma_2^z)$ at the mean position $(\bm{r}_1-\bm{r}_2)/2$ interacting with the electromagnetic vacuum. This emission of photons in a collective way known as super-radiance is effectively described in terms of collective angular momentum operators in the Dicke model \cite{Dicke}.

\textbf{Evaluation of the correlation measure.}
In order to numerically compute $\bar{I}$ for this dynamics, we consider a maximally entangled state $|\Phi_{\rmS\rmS'}\rangle$ between two sets $\rmS$ and $\rmS'$ of two qubits. Namely, $\rmS$ is the set of the two physical qubits, i.e. the two two-level atoms 1 and 2, and $\rmS'$ is made up of two auxiliary qubits $1'$ and $2'$ as sketched in Fig. 1. Next, the part $\rmS$ of the maximally entangled state $|\Phi_{\rmS\rmS'}\rangle\langle\Phi_{\rm \rmS\rmS'}|$ is evolved according to the master equation \eqref{MasterApp} while keeping the part ${\rmS'}$ constant, to obtain $\rho_{\rmS}^{\rm CJ}(t)$. This can be done, for instance, by numerically integrating the master equation $\frac{d \rho_{\rmS}^{\rm CJ}(t)}{dt}=\mathcal{L}\otimes\mathds{1} [\rho_{\rm S}^{\rm CJ}(t)]$, with the initial condition $\rho_{\rm S}^{\rm CJ}(0)=|\Psi_{\rmS\rmS'}\rangle\langle\Psi_{\rmS\rmS'}|$, where $\mathcal{L}$ is for the present example specified in Eq. \eqref{MasterApp}. Tracing out the qubits $2$ and $2'$ of $\rho_{\rmS}^{\rm CJ}(t)$ yields $\rho^{\rm CJ}_\rmS(t)|_{11'}$, and similarly tracing out qubits $1$ and $1'$ yields $\rho^{\rm CJ}_\rmS(t)|_{22'}$. Finally, this allows one to compute the von Neumann entropies of $\rho^{\rm CJ}_\rmS(t)|_{11'}$, $\rho^{\rm CJ}_\rmS(t)|_{22'}$ and $\rho_{\rmS}^{\rm CJ}(t)$ to calculate $\bar{I}(t)$ according to Eq. \eqref{Ibar}.

\end{document}